\documentclass[reprint,amsmath,amssymb,aps,groupedaddress,floatfix]{revtex4-1}
\usepackage{graphicx}
\usepackage{dcolumn}
\usepackage{bm}
\usepackage{natbib}
\usepackage{color,soul}
\usepackage{hyperref}

\begin{document}
\preprint{APS/123-QED}
\title{Transport phenomena in a free-standing two-dimensional sodium sheet}

\author{Ajit Jena}
 \altaffiliation[Also at ]{Key Laboratory of Optoelectronic Devices and Systems of Ministry of Education and Guangdong Province, College of Optoelectronic
Engineering, Shenzhen University, Shenzhen 518060, China}
\author{Wu Li}%
\affiliation{%
 Institute for Advanced Study, Shenzhen University, Shenzhen 518060, China \
}%
\date{\today}

\begin{abstract}
The advances in the growth techniques provide numerous scope to explore the possibilities of new 2D materials for potential applications. With the aid of first-principle calculations we show that 2D Na can be a new addition to the family of thermodynamically stable 2D materials for device applications. Not surprisingly, due to half-occupied $3s$ orbital 2D Na possesses the features of the 2D electron gas (2DEG). The transport properties are examined based on the accurate solution of Boltzmann transport equation. With practically tunable carrier density in 2D materials, the intrinsic electrical resistivity of electron doped 2D Na is $\sim$ 1.4 times larger than that of graphene and falls below the latter 450 K onwards. The Bloch-Gr\"uneisen temperature is almost constant at 50 K, independent of the type or density of the charge carriers. The electronic thermal conductivity of pure 2D Na is $\sim$ 1.24 times larger than that of its bulk counterpart at 300 K. The Wiedemann-Franz law stands tall in 2D Na with calculated Lorenz number 2.41 $\times 10^{-8} V^2/deg^2$ at room temperature. The transport mechanism presented here is expected to occur in all Na like systems with a clean Fermi surface.
\begin{description}
\item[Keywords]
2D Na, 2DEG, Bloch-Gr\"uneisen temperature, electron-phonon coupling, intrinsic electrical resistivity, Lorenz number 
\end{description}
\end{abstract}

\maketitle


\section{\label{sec:level1}introduction}
Two-dimensional (2D) materials have attracted considerable attention in the scientific community owing to their exceptional properties and promising applications. In this regard, graphene is at the forefront which has extremely large thermal conductivity and high mobile charge carriers having the dispersion curves similar to Dirac fermions with zero rest mass \cite{novoselov1, novoselov2, balandin, seol}. In addition to graphene the other 2D materials which also receive interest for new-generation electronic and spintronic devices include hexagonal boron nitride and members of the transition-metal chalcogenides such as molybdenum disulfide and tungsten diselenide \cite{butler, miro, zou}. There are also large number of literature which report the formation of 2D structures on a substrate or inside a material. For example, via scanning tunneling microscopy the existence of K islands ($\sim$ 5 - 500 nm) is found on graphite surface \cite{yin}. Superconductivity in one atomic layer metal film has been observed when Pb and In are grown epitaxially on Si(111) substrate \cite{zhang}. The fabrication of 2D crystalline layer of transition metal Hf on Ir(111) is also reported \cite{li}. A single atomic thick layer of Fe membrane is found to be suspended in graphene pores \cite{zhao}. In this context, theoretical work also predict that Ag, Au and Cu can be the stable 2D materials \cite{yang1, yang2, yang3}. The recent development of advanced growth techniques overcomes the challenge of identifying the new and potentially rewarding 2D materials which include phosphorene \cite{phsprn1, phsprn2} and materials for 2D ferromagnetic \cite{2dfm1, 2dfm2, 2dfm3}.   

Motivated by the intriguing physical properties of sodium and transition metal based layered oxides, we discuss the transport phenomena in a suspended sodium sheet. Sodium and transition metal based layered oxide, Na$_x$CO$_2$, offers a wide range of physical phenomena with different Na concentrations. For examples, Na$_{x}$CoO$_2$ shows interesting metal-insulator phase diagram, superconductivity is induced in Na$_{0.35}$CoO$_2$ when it is intercalated with water and Na$_x$Co$_2$O$_4$ is found to be a high thermoelectric power material \cite{foo, takada, wang}. All the above features make this hexagonal layered system an exciting transition metal oxide in which Na atoms lie in a plane. This encourages us to investigate the transport mechanism of a free-standing 2D Na. In a recent theoretical work, Nevalaita and Koskinen \cite{nevalaita} have studied 45 atomically thin elemental 2D metal films in hexagonal, square, and honeycomb lattice structures. They have predicted that 2D Na is mechanically stable in hexagonal and honeycomb lattices while unstable in the square lattice system. Since 2D Na is stable in the hexagonal lattice system we consider the same lattice type in the present study. In addition to Na we have also examined the dynamical stability (see the supplementary information) of Li, Be, Mg, Al and K which are reported to be mechanically stable 2D elements \cite{nevalaita}. However, we find that 2D Li and 2D Al have imaginary vibrational frequency in the phonon dispersion calculation (see the supplementary information). The absence of imaginary frequency in phonon dispersion and the energy evolution obtained from molecular dynamics (MD) run, shown in Fig.~\ref{fig:phdis-md}, suggest the thermodynamical stability of 2D Na.

In 2D metal, the carrier dynamics originating from electron-phonon (e-ph) interactions is characterized by Bloch-Gr\"uneisen temperature, $\Theta_{BG}$. $\Theta_{BG}$ is defined as $2\hbar k_{F} v_{s}/\kappa_{B}$, where $\hbar$, $k_{F}$, $v_s$ and $\kappa_{B}$ are respectively the reduced Planck constant, Fermi wavevector, sound velocity and the Boltzmann constant. $\Theta_{BG}$ is the temperature at which the resistivity starts deviating from the linear T behavior. Below $\Theta_{BG}$ the intrinsic electrical resistivity ($\rho_{e-ph}$) varies as $T^{4}$ while $\rho_{e-ph}$ is proportional to $T$ above $\Theta_{BG}$ \cite{fuhrer}. Efetov and Kim have shown that $\Theta_{BG}$ in graphene can be tuned up to $\sim$ 1000 K with high carrier density ($n = 4 \times 10^{14}$ cm$^{-2}$) by applying gate voltage \cite{gra-expt}. However, the effect of external carrier density on $\rho_{e-ph}$ in graphene is very negligible \cite{gra-expt, gra-res}. The scenario is significantly different in borophene, 2D B, where $\rho_{e-ph}$ is highly sensitive to the charge doping and $\Theta_{BG}$ is almost constant at 100 K \cite{borophene1}. In a very recent article Liu \textit{et al.} have shown that Bloch-Gr\"uneisen theory is not applicable to $\beta_{12}$ and $\gamma_{3}$ allotropes of borophene \cite{borophene2}. We note that, in the present discussion, $\Theta_{BG}$ in borophene refers to the temperature at which the resistivity starts showing non-linear T dependence on cooling. With electron and hole doping in 2D Na we show that $\Theta_{BG}$ is nearly pinned at 50 K and with electron doping $\rho_{e-ph}$ can be lower than that of graphene at high temperature (450 K onwards). The predicted results on the thermal transport in 2D Na are also promising when comparing with its bulk counterpart. We believe that 2D Na can be a new addition to the family of 2D materials for electronic and thermopower applications. The transport mechanism explained here is expected to be a prototype for the systems with clean Fermi surface and the work can also be useful to understand the transport phenomena in systems having a plane of Na atoms.

\section{\label{sec:level2}computational methodology}
Boltzmann Transport Equation (BTE) is known to describe well the electrical and thermal transport in insulators, semiconductors and metals \cite{allen, lindsay1, lindsay2, shengbte, wuli, epw}. We solve BTE accurately \cite{wuli} where the key component is the calculation of e-ph coupling matrix elements. We employ pseudo-potential based density-functional theory (DFT) and density-functional perturbation theory (DFPT) as implemented in Quantum ESPRESSO \cite{giannozzi} within the framework of general gradient approximation (GGA) to compute the electron energies, vibrational frequencies and e-ph matrix elements. The matrix elements are calculated first on a coarse grid of $8\times8\times1$ and then Wannier interpolated into a fine gird of $200\times200\times1$ using electron-phonon Wannier (EPW) package \cite{epw}. The calculations are performed using norm-conserving pseudo-potential and the kinetic energy cutoff for the planewave is taken as 60 Ry. The electronic integration over the Brillouin zone is approximated by the Gaussian smearing of 0.025 Ry for the self-consistent calculations. A single atom of Na is considered in the hexagonal lattice system ($a = 3.66$ \AA{}) and the Na sheets are sufficiently isolated from each other by 10 \AA{} of vacuum to ensure the negligible interlayer interaction. To carry out the MD simulation a supercell of $4\times4\times1$ is used.   

\section{\label{sec:level3}results and discussions}

\begin{figure}
\centering
\includegraphics[scale = 0.32]{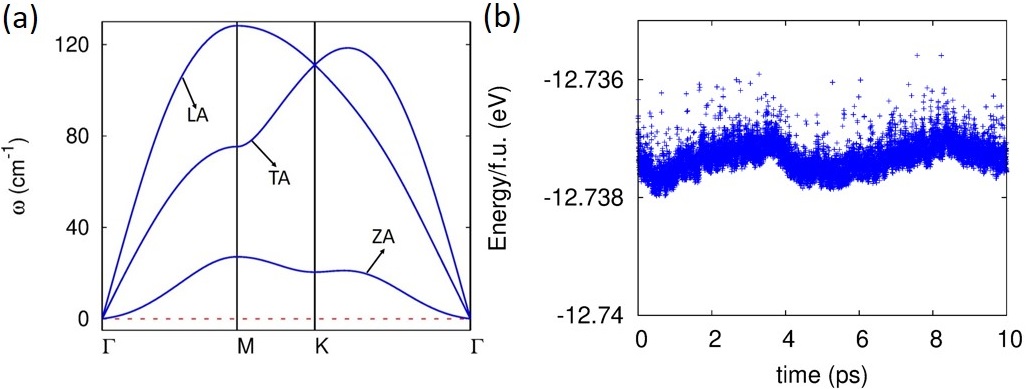}
\caption{(a) Phonon dispersion relation of optimized 2D Na along the high symmetry-lines of hexagonal lattice. (b) The room temperature energy evolution of 2D Na during MD simulation.}
\label{fig:phdis-md}
\end{figure}

Before we analyze the transport properties we present the phonon dispersion of 2D Na in Fig.~\ref{fig:phdis-md}(a). The absence of imaginary frequency in the dispersion curve suggests its structural satbility and thereby indicates that the experimental growth of the same can be plausible. The structural stability is further confirmed by the MD simulation carried out at 300 K. The change in the total energy is only within 2 meV ( see Fig.~\ref{fig:phdis-md}(b)) over the time steps. The phonon dispersion of this suspended sodium sheet comprises an out-of-plane (ZA) and two in-plane (LA and TA) modes. While the in-plane modes obey the normal linear dispersion around the $\Gamma$-point the soft ZA mode shows $q^{2}$ frequency dispersion, typical in 2D materials, which is a consequence of $D_{6h}$ point group \cite{d6h}. As it can be seen from Fig.~\ref{fig:phdis-md}(a) the highest phonon energy ($\sim$ 130 cm$^{-1}$) in 2D Na is significantly lower than (by $\sim$ 1 order) that of graphene and borophene \cite{gra-phonon, borophene1, borophene2}. We note that the highest phonon energy in bulk Na \cite{bulk-phonon1, bulk-phonon2} is of similar magnitude to that of 2D Na. Therefore, the Debye temperature ($\Theta_{D}$) is lower in bulk Na as compared to B and C \cite{borron-thetaD, kittel1}. $\Theta_{D}$ is the temperature that corresponds to the highest normal mode of vibration in crystal. $\Theta_{D}$ in bulk systems is equivalent to $\Theta_{BG}$ in 2D, systems with low electron density, while describing $\rho_{e-ph}$. The high temperature behavior of $\rho_{e-ph}$ in typical conductors (3D) is same as in 2D ($\propto T$). However, $\rho_{e-ph}$ varies as $T^{5}$ below $\Theta_{D}$ in 3D systems. Due to large Fermi surface in most of the metals all phonons are able to scatter electrons \cite{fuhrer} implying $\Theta_{BG}$ is equal to $\Theta_{D}$. 

Charge carrier density is considered as an important parameter to engineer the physical properties in 2D materials. For example, with increasing carrier density MoS$_2$ exhibits metal-insulator transition whereas a medium carrier density is needed for its transistor application \cite{transistor-mos2}. It is reported that by hole doping ($n = +~3.3 \times 10^{14}$ cm$^{-2}$), $\rho_{e-ph}$ of pristine $\beta_{12}$ borophene increases by $\sim$ 4.29 times \cite{borophene1}. And as mentioned earlier, $\Theta_{BG}$ of graphene can be varied from 100 to 1000 K with different carrier densities \cite{gra-expt}. While the effect in the former example is due to strong electron-electron interaction, in the latter two cases it is limited by the electron-phonon coupling. Since the 2D metal surface is quite exposed to external gate the carrier density can be tuned considerably by applying gate voltage. Concurrently, the Fermi energy (E$_F$) of the system changes with gate voltage. Park \textit{et al.} have shown that the E$_F$ in graphene is found to lie at $\sim$ 0.65 eV from the Dirac point corresponding to $4 \times 10^{13}$ cm$^{-2}$ of the carrier density \cite{park}. In borophene, 0.43 eV of shifting of E$_F$ leads to the carrier density $n = 3.3 \times 10^{14}$ cm$^{-2}$ \cite{borophene1}. The preexisting metallic nature of pristine borophene \cite{borophene1, borophene2} is attributed to this large carrier density with small change in E$_F$.  

\begin{figure}
\centering
\vspace*{-0.4cm}
\includegraphics[scale = 0.35]{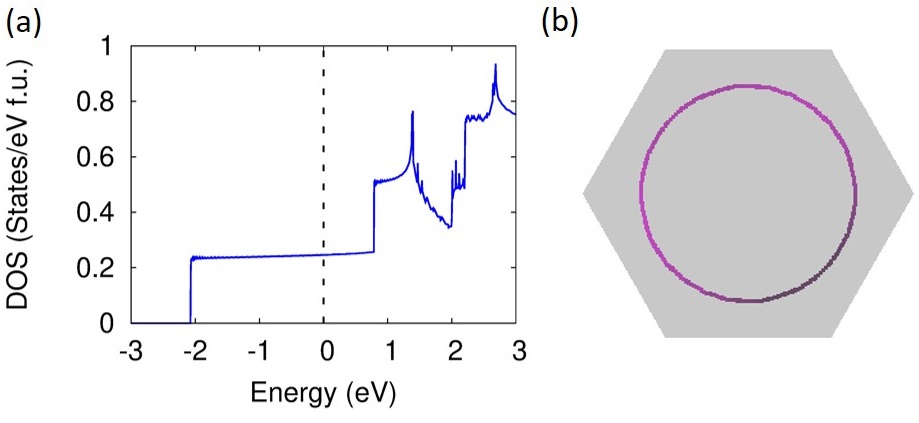}
\caption{(a) Density of states (DOS) of 2D Na. The Fermi energy (E$_F$) is set to zero. The DOS around E$_F$ is nearly constant, shows the typical nature of 2DEG. (b) Fermi surface of 2D Na in the first Brillouin zone.}
\label{fig:dos}
\end{figure}

\begin{figure}
\centering
\vspace*{0.2cm}
\includegraphics[scale = 0.5]{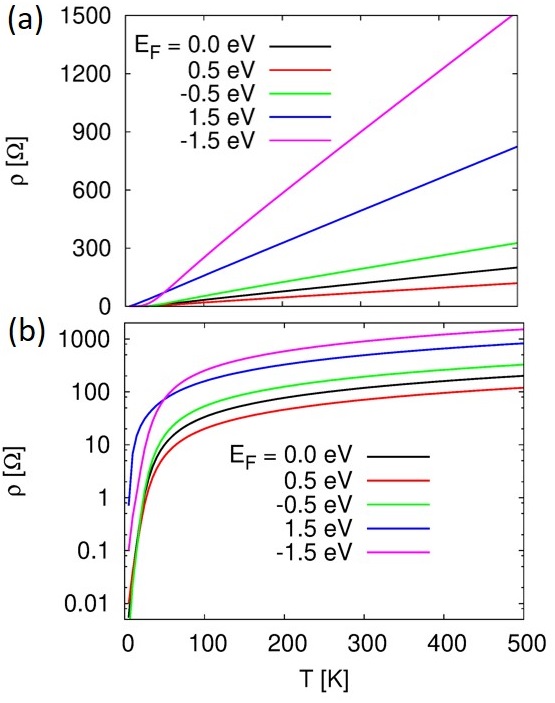}
\caption{(a) Temperature dependence intrinsic electrical resistivity of 2D Na for different shifting of E$_F$. (b) Similar quantities as in (a) plotted in logarithm scale to distinguish the behavior of $\rho_{e-ph}$ in two distinct temperature regimes.}
\label{fig:res-log}
\end{figure}

To incorporate the charge doping we have shifted the E$_F$ with respect to the E$_F$ of pristine 2D Na. Then from density of states (DOS) we have estimated the carrier density. At E$_F$ = 0.5 eV the carrier density is found to be $2.23 \times 10^{14}$ cm$^{-2}$. Achieving such a carrier density is feasible in 2D materials by external gate voltage \cite{gra-expt}. The DOS of 2D Na is shown in Fig.~\ref{fig:dos}(a) which explains the typical nature of 2DEG, DOS is independent of energy. Therefore, we have also calculated the electron effective mass by using the relation in 2DEG, $n = 2m^\ast E_F/\pi \hbar^2$, where $m^\ast$ is effective mass of the electron. For E$_F$ = 0.5 eV, we get $m^\ast$ as 0.53 $m_e$ with $m_e$ being the mass of electron. This is nearly equal to the effective mass of 0.47 $m_e$ that we have calculated from the $E(k)$-$k$ dispersion. The characteristic features of electron-phonon interaction driven resistivity of 2D Na are shown in Fig.~\ref{fig:res-log}. Fig.~\ref{fig:res-log}(a) represents $\rho_{e-ph}$(T) for different E$_F$. For pure case E$_F$ is considered as 0 eV. $E_F > 0$ $(E_F < 0)$ represents electron (hole) doping. The results suggest that $\rho_{e-ph}$ is largely dependent on E$_F$. For E$_F$ = -0.5, 1.5 and -1.5 eV, $\rho_{e-ph}$ is respectively $\sim$1.5, $\sim$4.0 and $\sim$7.0 times larger than the $\rho_{e-ph}$ (at T = 500 K) of pristine 2D Na. However with E$_F$ = 0.5 eV, $\rho_{e-ph}$(T) lies below the $\rho_{e-ph}$(T) of pure compound. The mechanism behind this reduced resistivity is discussed later. We believe it is one of the important predictions in this new 2D material. 

\begin{figure}
\centering
\vspace*{0.2cm}
\includegraphics[scale = 0.31]{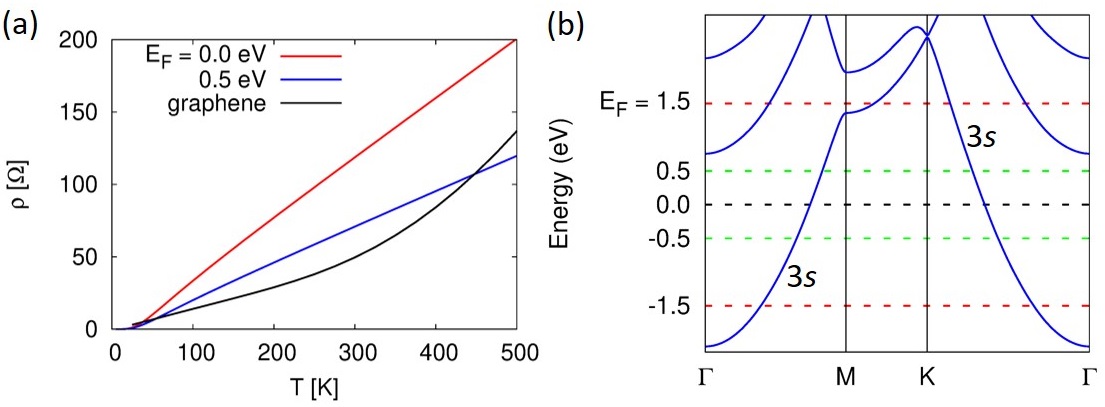}
\caption{(a) Temperature dependence intrinsic electrical resistivity of 2D Na in comparison with graphene. (b) Band structure of 2D Na along the high-symmetry lines of hexagonal lattice. The E$_F$ is 0.0 eV for the pure case.}
\label{fig:gra-res-band}
\end{figure}

The carrier density that is obtained from E$_F$ = $\pm 1.5$ eV is yet to be achieved by experiment in 2D materials. We note that E$_F$ = $\pm 1.5$ eV are the two hypothetical case studies carried out for comparison purpose only. At E$_F$ =  1.5 eV, we have two bands across the Fermi energy (see Fig.~\ref{fig:gra-res-band}(b)). Therefore, the transport mechanism at E$_F$ = 1.5 eV can not be simple like at E$_F$ = 0.0 and 0.5 eV. In fact, due to intricate Fermi surface (FS) Bloch-Gr\"uneisen theory is not satisfied in $\beta_{12}$ and $\gamma_{3}$ borophene \cite{borophene2}. In contrast to borophene, the FS in 2D Na is clean up to E$_F$ $\approx$ 0.79 eV. Fig.~\ref{fig:dos}(b) represents the FS of 2D Na at E$_F$ = 0.0 eV. The circular shape of the FS holds another characteristic of 2DEG. Though it is not shown here the FS at 0.5 eV is also a circle. Up to E$_F$ $\approx$ 0.79 eV, the electronic conduction can be described by a single band ($3s$). Hence, the associated FS is isotropic which indicates the applicability of Bloch-Gr\"uneisen theory. Fig.~\ref{fig:res-log}(b) represents $\rho_{e-ph}$(T) in the logarithm scale for various E$_F$. Two distinct regimes of $\rho_{e-ph}$(T) are well noticed from the plot. The crossover between the two regimes occurs at temperature , $\Theta_{BG}$, $\sim$ 50 K and is independent of E$_F$ and $n$ ($n$ is proportional to E$_F$). The phonons are excited at low temperature to scatter the carriers in this soft-mode system. The carrier density dependence behavior of $\Theta_{BG}$ in 2D Na is similar to the borophene case and is different from graphene.\\

\begin{figure}
\centering
\includegraphics[scale = 0.55]{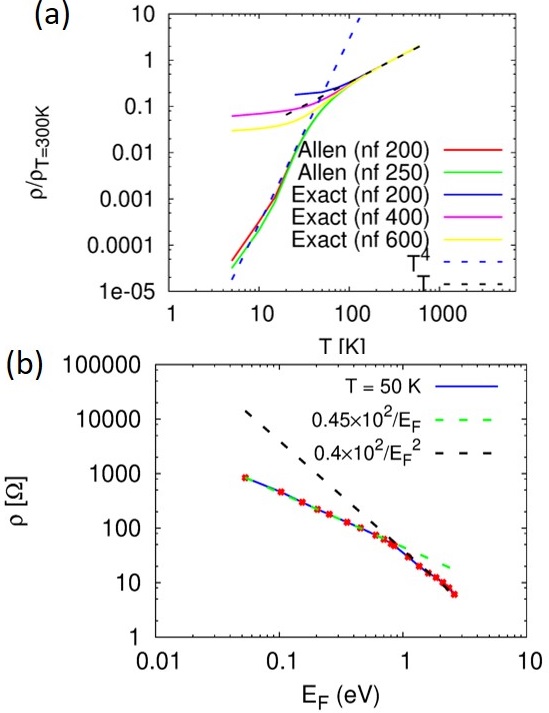}
\caption{(a) Normalised electrical resistivity, in the double log scale, of 2D Na calculated using Allen's model and an exact solution with different grids.The dashed blue and dashed black lines are fitted by the functions $\rho \sim T^4$ and $\rho \sim T$ respectively. (b) Variation of $\rho$ with E$_F$. The E$_F$ values are w.r.t. the band minimum and the latter is taken as origin}
\label{fig:double-log}
\end{figure}

The $\rho_{e-ph}$(T) of 2D Na that we discussed above are calculated with Allen's model \cite{allen}. We have also calculated $\rho_{e-ph}$(T) with an exact solution \cite{wuli}. We find that the results are in good agreement in both the calculations. However, at very low temperature Allen's model explains accurately the $T^{4}$ behavior of 2D Na as shown in Fig.~\ref{fig:double-log}(a). To demonstrate the low and high temperature behaviors of resistivity we have shown the normalised resistivity ($\rho$/$\rho_{T=300K}$), in the double log scale, calculated using the aforementioned approximations with different grids. The gird of $200\times200\times1$ is represented by ``nf 200" in the figure. We have fitted $\rho$/$\rho_{T=300K}$ to $A T^4$ and $B T$ respectively for the low and high temperature regimes with $A = 2.9 \times 10^{-8}/K^4$ and $B = 3.3 \times 10^{-3}/K$. The low and high temperature fitted lines are intersected at $\Theta_{BG}$ ($\sim$ 50 K). As we have seen from Fig.~\ref{fig:res-log}(a), besides the case of E$_F$ = 1.5 eV, the resistivity increases monotonically with E$_F$. At E$_F$ = 1.5 eV, we have another band across the Fermi energy (see Fig.~\ref{fig:gra-res-band}(b)). We have included more studies to see the behavior of resistivity with E$_F$ (see the supplementary information) for which the band minimum is taken as origin. The power law dependence of $\rho$ on E$_F$ is explained through Fig.~\ref{fig:double-log}(b) for a particular temperature (= 50K). The resistivity varies as 1/E$_F$ and 1/E$_{F}^{2}$ respectively for the lower and higher E$_F$. This can be understood through the relation, $\rho = 2/(e^2 v_{F}^{2} D(E_F)\tau(E_F))$, where $v_F$ ($\propto \sqrt{E_F})$ is the Fermi velocity and $\tau$ is the carrier life time \cite{sarma-rmp}. Near the band edge i.e. for lower E$_F$, the scattering rate (1/$\tau$) mimics the DOS of 2D Na, constant of energy. Relation between the scattering rates and DOS near the band edge has already been discussed in references \cite{wuli, lundstrom}. The scattering rates at an energy away from the band edge decreases with energy where DOS is still independent of energy. The scattering rates at lower and higher energies are provided in the supplementary information for different temperatures. This suggests that the resistivity varies as 1/E$_F$ and 1/E$_{F}^{2}$ respectively in the lower E$_F$ and higher E$_F$. This also implies that $\rho$ is proportional to 1/$n$ (1/$n^{2}$) at low (high) electron density ($n$ $\propto$ E$_F$). The electrical transport in bilayer graphene, also a system with parabolic band dispersion, has similar behavior \cite{sarma-rmp}. In the following paragraph we discuss the conductive behavior for a specific carrier density that can be achieved in practice (E$_F$ = 0.5 eV).                              

Since we have enhanced conductivity for E$_F$ = 0.5 eV in 2D Na it is important to compare the values quantitatively with the 2D material with least known electrical resistivity, graphene. Through Fig.~\ref{fig:gra-res-band}(a) we have presented the phonon limited temperature dependence electrical resistivity of graphene with pure and doped 2D Na. The $\rho_{e-ph}$ of electron doped (E$_F$ = 0.5 eV) 2D Na is about 1.4 times larger than the $\rho_{e-ph}$ of graphene. But, interestingly the former falls below the latter 450 K onwards. The phonons in graphene are excited around this temperature to scatter the carriers and possesses the increased resistivity. To understand the improved conductivity in doped 2D Na, compared to the pure case, we have calculated the slope ($\propto$ velocity) of the $3s$ energy band in both cases. We find that the slope around E$_F$ = 0.5 eV is $\sim$ 1.3 times larger than around 0.0 eV. This suggets that the $3s$ band is more dispersive at 0.5 eV than at 0.0 eV. More dispersion leads to smaller effective mass and hence the electron mobility increases. A flat band possesses large effective mass with weak conductivity. In Fig.~\ref{fig:gra-res-band}(b) we have labelled different Fermi energies in the electronic band structure of 2D Na. A careful observation of Fig.~\ref{fig:gra-res-band}(b) tells us that the Fermi energy is close to the flat regime of the $3s$ orbital at large hole doping (E$_F$ = - 1.5 eV). This leads to substantial increase in the resistivity values (see Fig.~\ref{fig:res-log}(a)). Therefore, pinning E$_F$ at a particular energy one can manipulate the conductive behavior in 2D Na. It is expected that all Na like systems will have similar kind of transport mechanism as explained above.               

\begin{figure}
\centering
\vspace*{0.4cm}
\includegraphics[scale = 0.5]{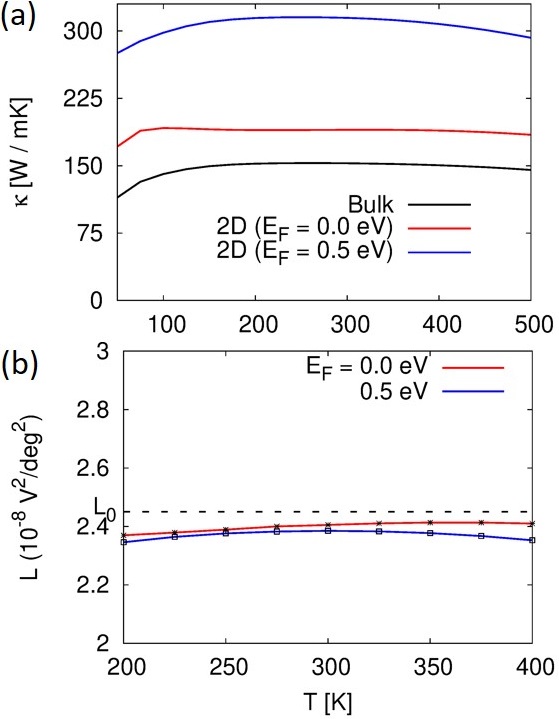}
\caption{(a) Temperature dependence electronic thermal conductivity of bulk and pure and doped 2D Na. (b) Comparison of Lorenz number in bulk and pure 2D Na.}
\label{fig:kappa}
\end{figure}

We now examine the thermal transport and the validity of Wiedemann-Franz law in 2D Na. In metals, the total thermal conductivity $\kappa$ is defined as $\kappa_{e}$ + $\kappa_{ph}$, where $\kappa_e$ is contributed by the electrons and $\kappa_{ph}$ is due to phonons. However in free-electron-like systems (e.g. Na and K), $\kappa_{ph}$ is negligible and $\kappa$ is mainly contributed by the electrons. We also calculate $\kappa_{ph}$ by modifying ShengBTE package \cite{shengbte} and find that $\kappa_{e}/\kappa_{ph} \approx 8.8 \times 10^{4}$ at 300 K in 2D Na. Therefore, $\kappa_{ph}$ is not considered in the further discussion and we are interested only on $\kappa_e$. In typical metals, $\kappa_e$ is $\propto$ T in the low temperature limit and at high temperature $\kappa_e$ is independent of T. Fig.~\ref{fig:kappa}(a) shows the temperature dependence features of $\kappa_e$ for pure and electron doped (E$_F$ = 0.5 eV) 2D Na in comparison with bulk Na. The figure demonstrates that the theory of electronic thermal transport is not violated both in low and high temperature limits. Moreover, we find that $\kappa_e$ of pure 2D Na is $\sim$ 1.24 times larger than the $\kappa_e$ of bulk Na at room temperature. It is due to the reason that 2D Na has increased electron life time. The scattering rates of the bulk and 2D Na are compared in the supplementary information. 

Additionally, it is found that $\kappa_{e}^\text{doped-2DNa}$ $>$ $\kappa_{e}^\text{pure-2DNa}$, where $\kappa_{e}^\text{doped-2DNa}$ stands for $\kappa_{e}$ of doped 2D Na. Previously we have seen that $\rho_{e-ph}^\text{doped-2DNa}$ ($E_F$ = 0.5 eV) $<$ $\rho_{e-ph}^\text{pure-2DNa}$. Hence, $\kappa_{e}^\text{doped-2DNa}$ ($E_F$ = 0.5 eV) $>$ $\kappa_{e}^\text{pure-2DNa}$ can only happen when the ratio of the thermal conductivity ($\kappa$) to the electrical conductivity ($\sigma$) is constant. For metals, at not too low temperature $\kappa/\sigma$ is directly proportional to the temperature and is defined as $\kappa/\sigma = LT$, where $L$ ($= \pi^2 \kappa_{B}^{2}/3e^2$) is termed as the Lorenz number \cite{kittel2}. The Lorenz number does not depend on the scattering mechanisms or on the dimensionality of the system. The calculated $L$ combinedly for pure and doped 2D Na lies in the range 2.35 - 2.41 $\times 10^{-8} V^2/deg^2$ (see Fig.~\ref{fig:kappa}(b)) which is not much deviated from the theoretical value, $L_0$ (= 2.45 $\times 10^{-8} V^2/deg^2$). In the supplementary information we have included the derivation of $L$ both for 2D and 3D metals. The derivation is based on the postulate that the free electrons are the primary carriers both for charge and heat current.      


\section{\label{sec:level4}summary and conclusions}
In summary, we carry out \textit{ab initio} calculations to investigate the electrical and thermal transport of a free-standing two-dimensional (2D) sodium sheet based on the accurate solution of Boltzmann transport equation. The results suggest that 2D Na behaves like 2DEG where the electronic conduction is primarily driven by the half-filled $3s$ orbital. With achievable carrier density in 2D systems, we find that the temperature dependence intrinsic electrical resistivity of electron doped 2D Na is $\sim$ 1.4 times larger than that of graphene and lies below than that of the latter 450 K onwards. Bloch-Gr\"uneisen temperature is predicted to lie at $\sim$ 50 K in this soft-phonon-mode system and is not dependent on the type or concentration of the charge carriers. The electronic thermal conductivity ($\kappa_e$) of pure 2D Na is $\sim$ 1.24 times larger than the $\kappa_e$ of bulk Na at 300 K. The Wiedemann-Franz law is not violated in 2D Na. The results presented here are not only encouraging from the 2D electronic and thermopower devices viewpoint but are also important for the bulk systems having a plane of Na atoms.  

\section*{Acknowledgements}
A. Jena acknowledges the financial support from Shenzhen Science, Technology and Innovation Commission.

\bibliography{paper} 
\providecommand{\noopsort}[1]{}\providecommand{\singleletter}[1]{#1}%
\end{document}